\documentclass[fleqn,usenatbib]{mnras}

\usepackage{newtxtext,newtxmath}

\usepackage[T1]{fontenc}

\DeclareRobustCommand{\VAN}[3]{#2}
\let\VANthebibliography\thebibliography
\def\thebibliography{\DeclareRobustCommand{\VAN}[3]{##3}\VANthebibliography}

\usepackage{graphicx}	
\usepackage{amsmath}	

\title[Some statistical remarks on GRBs jointly detected by Fermi  and Swift satellites]{Some statistical remarks on GRBs jointly detected by Fermi  and Swift satellites
}

\author[S. Pinter et al.]{
Sandor Pinter$^{1,2}$\thanks{E-mail: sandor.pinter@uni-nke.hu},
Lajos G. Balazs$^{2,3}$,
Zsolt Bagoly$^{4,1}$,
L. Viktor Toth$^{2}$,
Istvan I. Racz$^{1}$,
and Istvan Horvath$^{1}$
\\
$^{1}$University of Public Service, Budapest, Hungary\\
$^{2}$Department of Astronomy, E\"otv\"os University, Budapest, Hungary\\
$^{3}$Konkoly Observatory, Research Centre for Astronomy and Earth Sciences, Budapest, Hungary\\
$^{4}$Department of Physics of Complex Systems, E\"otv\"os University, Budapest, Hungary
}

\date{Accepted XXx. Received YYy; in original form ZZZ}

\pubyear{2023}

\begin{document}
\label{firstpage}
\pagerange{\pageref{firstpage}--\pageref{lastpage}}
\maketitle

\begin{abstract}
We made statistical analysis of the Fermi GBM and Swift BAT observational material, accumulated over 15 years. We studied how GRB parameters (T$_{90}$ duration, fluence, peak flux) that were observed by only one satellite differ from those observed by both. In the latter case, it was possible to directly compare the values of the parameters that both satellites measured. The GRBs measured by both satellites were identified using the \textit{knn()} k-nearest neighbour algorithm in the \textit{FNN} library of the \textbf{R} statistical package. In the parameter space we determined the direction in which the jointly detected GRBs differ most from those detected by only one of the instruments using the \textit{lda()} in \textit{MASS} library of \textbf{R}. To get the strength of the relationship between the parameters obtained from the GBM and BAT, a canonical correlation was performed using the \textit{cc()} procedure in the \textit{CCA} library in \textbf{R}. The GBM and BAT T$_{90}$ distributions were fitted with a linear combination of lognormal functions. The optimal number of such functions required for fit is two for GBM and three for BAT. Contrary to the widely accepted view, we found that the number of lognormal functions required for fitting the observed distribution of GRB durations does not allow us to deduce the number of central engine types responsible for GRBs.
\end{abstract}




\section{Introduction}

GRBs have been known for decades since their discovery by \citet{1973ApJ...182L..85K}. Nevertheless, there is still no generally accepted theory for their origin, which would fully and satisfactorily explain all the observational facts. The first measurements of the BATSE instrument on board the CGRO satellite \citet{1992NASCP3137...26F} have already shown that there are two characteristic maxima in the distribution of the T$_{90}$ duration: one is in the $0.1-1$s, the other is in the $10-100$s time frame \citep{1993ApJ...413L.101K}. In our case, as we see later, these two characteristic peaks are in both Fermi and Swift T$_{90}$ distributions, although the times for the peaks are different. In the case of short bursts this peak is at 0.6s for Fermi and at 0.3s for Swift.

To explain these two peaks, researchers generally agree that the models can be divided into two large groups. One of them assumes that GRBs are originating from merging two compact objects (neutron star, black hole, or possibly a white dwarf). Analyzing GRB light curves one can find those that best fits one of these mechanisms \citep{2018JCAP...10..006R}.

GRBs with long ($> 10$s) durations are typically caused by collapsing high mass (> 10 solar mass) stars. In some cases, however, merging two compact objects may also produce long-GRBs \citep{2018EPJWC.16801006R}. \citet{2007MNRAS.374L..34K} also presented arguments for producing long GRBs by merging a massive white dwarf with a neutron star. Of course, all these models are theoretical possibilities.

There are also ideas that cannot be fitted to any of the above models. \citet{2003ApJ...594..919H}, for example, believe that GRBs may also be formed by a neutron star kick. \citet{2000ApJ...530L..69B} studied the conversion of a neutron star to a strange star as a possible energy source for GRBs.

Naturally, all these models can be realized, however, not necessarily with the same frequency in a GRB sample, collected from observations. The two well-defined peaks in the T$_{90}$ distribution may indicate that one of the models is dominant for the short and long GRBs, respectively. Of course, this is just a statistical argument. In the case of some specific GRBs, it is necessary to carefully analyze whether one of the options has been realized or whether we are facing a new case that has not been studied in theory so far.

The distribution of T$_{90}$, observed by BATSE, could be approximated by the superposition of two lognormal distributions. 
However, \citet{1998ApJ...508..757H} and \citet{1998ApJ...508..314M} showed that supposing a third, intermediate lognormal group fits much better the T$_{90}$ distribution.
Many authors 
\citep{hak00,bala01,hor02,bor04,hor04,chat07,zito15} have since confirmed the existence of this Intermediate 
GRB class in the same database
using different techniques.

Analyzing T$_{90}$ distribution obtained by the Swift satellite also resulted in the existence of a third, intermediate group between the short and long GRBs \citep{hor08,huja09,hor10,zito15,ht16,2022ApJDeng}. 
For the Fermi GRBs there are also many single or multi dimensional analysis, about this topic
\citep{2015A&A...581A..29T,tarnopolski2016,hoi19ApS,2019ApJ...870..105T,2022MNRAS517.5770Z,2022GalaxSalmon}.

Whether we look at merging or collapsar models, a very compact object is created for both types. This is the fireball model by \citet{1993ApJ...405..278M}. The energy condensed in this extremely small volume is released in a very short-lived explosion and creates the GRB phenomenon observed    \citep{2004RvMP...76.1143P,2006RPPh...69.2259M,2015AdAst2015E..22P}.

The compact objects created in the models outlined above, in which the compressed energy is released in the form of GRB, differ in the extremely small volume in the compressed energy and in the time scale of the burst dynamics. The lognormal peak in the T$_{90}$ distribution supports the dominance of any of these.

The question arises, does the third lognormal peak suggest the presence of a third type of central engine for intermediate T$_{90}$ duration GRBs? We get closer to the answer, if we look at the GRBs that both Fermi and Swift detected.

\subsection{Differences in observations' strategies}

The Neil Gehrels Swift Observatory and the Fermi Gamma-ray Space Telescope have different technical layout and observational strategy. Swift has three major observational facilities: a coded mask for gamma-ray detection (Burst Alert Telescope, BAT), and two telescopes for X-ray and  Ultra-Violet/Optical range (XRT and UVOT, respectively) \citep{2004ApJ...611.1005G,2005SSRv..120..143B}.

The Swift is operating in observatory mode, which means, after getting a burst alert, the BAT is slewing to point to the burst’s direction in the sky \citep{2000AIPC..526..731B,2005SSRv..120..143B}. BAT covers a large fraction of the sky (over one steradian fully coded, three steradians partially coded; by comparison, the full sky solid angle is $4\pi$ or about 12.6 steradians). It locates the position of each event with an accuracy of 1 to 4 arc-minutes within 15 seconds. The BAT is sensitive in the $15-150$ keV energy range.

The XRT can take images and perform spectral analysis of the GRB afterglow. This provides more precise location of the GRB, with a typical error circle of approximately 2 arcseconds radius. The XRT is also used to perform long-term monitoring of GRB afterglow light-curves for days to weeks after the event, depending on the brightness of the afterglow \citep{2000SPIE.4140...64B,2000SPIE.4140...87H,1996SPIE.2805...56C,1996SPIE.2808..414H,1998SPIE.3445...13S,1992SPIE.1546..205W,1997SPIE.3114..392W}. The XRT is sensitive in the $0.2-10$ keV energy range.

The UVOT is used to detect optical afterglows. The UVOT provides a sub-arcsecond position and makes optical and ultra-violet photometry \citep{2005SSRv..120...95R,2001A&A...365L..36M,1989MNRAS.237..513F}.

The Swift strategy is to reach all new GRB positions as soon as possible and follow all the GRB afterglows as long as the signal can be distinguished from the background noise of the detector. The rotation time of the Swift baseline is less than about 90 seconds. XRT and UVOT observations begin while the burst is still in progress. When Swift is blocked in pointing to prompt observations of the most recent bursts, it will follow a schedule uploaded from the ground. This schedule makes possible to follow-up the GRB afterglows when they are in the line of sight of the detectors as long as possible, until the observable brightness of the burst become fainter than the sensitivity threshold of the detector.

Fermi includes two scientific instruments, the Large Area Telescope (LAT) and the Gamma-ray Burst Monitor (GBM) \citep{fermi1,fermi2,fermi3,fermi4,fermi5,von_Kienlin_2020}. 
The LAT is an imaging gamma-ray detector (a pair-conversion instrument) which detects photons with energy from about 20 MeV to 300 GeV, with a field of view of about 2.5 steradian (20\% of the whole sky) \citep{fermi_lat_technical}.

The GBM consists of 14 scintillation detectors (twelve sodium iodide crystals for the 8 keV to 1 MeV range and two bismuth germanate crystals with sensitivity from 150 keV to 30 MeV), and can detect gamma-ray bursts in that energy range across the whole $4\pi$ area of the sky not occluded by the Earth \citep{fermi_gbm1,fermi_gbm2}.

For the first few years of the Fermi mission the default observation mode was an all sky survey, optimized to provide relatively uniform coverage of the entire sky with the LAT instrument every three hours. More than 95\% of the missions were carried out in this observation mode. However, Fermi’s flexible survey mode is capable of patterns and inertially pointed observations, all of which allow for increased coverage of selected parts of the sky. 

Due to the different energy response characteristics, technical layout, and observational strategy, the GRBs detected by Swift is not necessarily detected by Fermi and vice versa. 
It is an important problem, therefore, to study which part of the GRB population is observed by both of the satellites and which one is observed only by one of them. Furthermore, it is also important to know if there are physical differences between these classes \citep{racz_mnras_fermi_2018,racz_an_fermi_2018}. 

To compare the physical parameters of the GRB detected by BAT and GBM, we used the physical quantities obtained from measurements of both satellites. These parameters are the following: duration (T$_{90}$), fluence, 1024 ms peak flux.

\subsection{Comparison of BAT and GBM GRB triggering}

The BAT burst trigger algorithm looks for count rates over the estimated background and constant sources. The algorithm is constantly examining the criteria that determine the preburst background. The BAT processor continuously follows hundreds of such criteria in the same time. The eruption trigger threshold is adjustable by program between $4-11$ sigma above background noise, typically 8 sigma value. One of the most important features of BAT is its imaging ability. 

After the burst triggers, the onboard software checks that the trigger comes from a point source, thus many background sources can be eliminated. This yields a GRB fluence sensitivity of $\approx\,10^{-8}\, erg\, cm^{-2}\, s^{-1}$ (in 15-150 keV rage), corresponding to $\approx\,0.1\, cm^{-2}\, s^{-1}$ at 75 keV, the middle of the energy range of BAT sensitivity. 

A GBM burst trigger occurs when the onboard software detects an increase in the count rate of two or more NaI detectors above an adjustable threshold in units of background count rate standard deviation ($4.5-7.5\sigma$). The trigger algorithms uses four BATSE compatible energy ranges ($25-50$~keV, $50-300$~keV, $100-300$~keV, and $>300$keV) and ten different timescales between 16ms---8.192s. There are 120 distinct trigger algorithms available, with approximately 75 of them typically operating concurrently.
Fermi GBM's burst sensitivity (the peak $50-300$ keV flux for $5\sigma$ detection)  is less than $0.5 ph \, cm^{-2} \, s^{-1}$. 

Hence, the background estimation is different for the two satellites and therefore the integrated $T_90$ calculation methods are using different methods. The BAT's coded mask restoration algorithm inherently includes the background subtraction, leaving only the statistical fluctuation in the lightcurve data. In the GBM case the values of the $T_90$ from a lightcurve relies on the background estimation. To estimate the background during a GBM trigger, a common technique is to select background intervals on either side of the trigger and interpolate using a polynomial function. Another approach involves acquiring background spectra from orbits on preceding and subsequent days when the spacecraft occupied a similar geomagnetic position in its orbit \citep{2012SPIE.8443E..3BF}. For the precise background determination one could also take the detailed
positional information of the satellite and the celestial objects (Earth, Sun, Moon) into account \citep{2013A&A...557A...8S}, or use a  physically motivated detailed background model for the GBM \citep{2020A&A...640A...8B}. Other information maximalization techniques can alse be used, e.g. the Automatized Detector Weight Optimization which maximizes the signal's peak over the background's peak over the search interval \citep{2016A&A...593L..10B}.  

Although, the energy sensitivity range of GBM is much wider than that of BAT, the higher sensitivity of BAT might resulted in triggering GRBs in BAT but not in GBM. It can happen that only one of BAT or GBM is triggered, but if it is the case at both satellites, the observed physical parameters of GRBs will be different due to the different spectral characteristics of BAT and GBM.

\section{Data \& Methods}

Our main database consists all GRBs of Swift\footnote{Swift BAT: https://swift.gsfc.nasa.gov/archive/grb\_table/} and Fermi\footnote{Fermi GBM: https://heasarc.gsfc.nasa.gov/W3Browse/fermi/fermigbrst.html} detections from the beginning of their missions (December 17$^{th}$, 2004 for Swift and July 14$^{th}$, 2008 for Fermi) until April 14$^{th}$, 2023. 
We used only those GRBs in our analysis when both satellites were observing simultaneously: from the first observation of Fermi. 

\begin{figure}
    \centering
    \includegraphics[width=0.7\columnwidth]{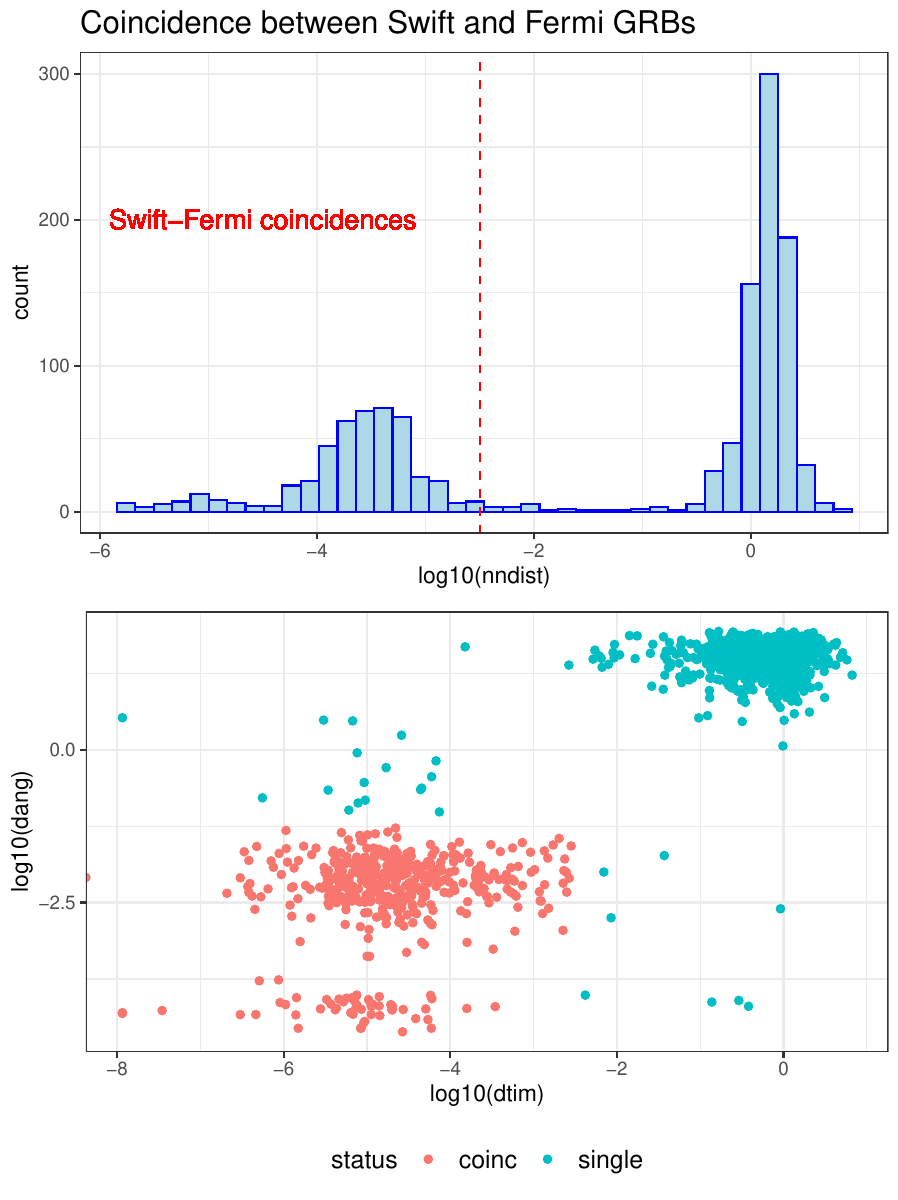}
    \caption{Upper panel shows frequency distribution of Euclidean nearest neighbor distances between Fermi and Swift GRBs, in angular position-trigger time parameter space. 
     Red dashed line marks the boundary between real and random coincidences. Lower panel shows the distribution of nearest neighbor GRBs in the angular  position (measured in degrees) trigger time difference (measured in days) plane.  Light red points (status coinc) indicate real coincidences.
     The small lower bump in the left of the image represent GRBs not having durations estimated independently in BAT and GBM data.}
    \label{fig:nn-dist}
\end{figure}

First we assigned an angular position-trigger time frame to the GRBs detected by the Swift and Fermi satellites, respectively. For the detailed procedure see \citet{racz_an_fermi_2018,racz_mnras_fermi_2018}. Then we identified the closest Fermi-Swift pairs in this coordinate frame using the $knn$ procedure in the $FNN$ library of the $\bf R$ statistical program \citep{R,FNN,ripley_1996,venables_ripley_2002}. The results can be seen in Fig. \ref{fig:nn-dist}. 

\subsection{Comparing the physical properties of "couples" and "widows" GRBs in BAT and GBM}

We have already mentioned in the introduction that the technical layout of the Swift and Fermi satellites and, consequently, their observational strategies are different. So the question arises on which it depends whether a burst is detected by both satellites, and when only one of them. It may be a simple geometric effect, i.e. the corresponding burst is not in the observed region of the sky at one of the satellites. 

In that case if the burst is in the field of view of both satellites but below the detection limit of one of them, the statistical distribution of the physical parameters of the bursts could be different. Namely, in case of a simple geometric selection effect the statistics of the physical properties of both the "couples" and "widows" bursts should be the same. In the second case, however, when the successful observation depends on the detection limit of the instrument it is not necessarily true. 
 
Motivated by these facts it is worth comparing the statistical properties  of "widows" and "couples" GRBs detected by both or only one of the satellites. In the following we discuss these issues in case of Swift BAT and Fermi GBM, separately.

\subsection{Creating "couples" and "widows" frames}

The results of computing $k$-nearest neighbour distances enabled us to create three data frames: Swift-Fermi "couples", Swift "widows", and Fermi "widows". These names refer to GRBs detected by both Swift and Fermi satellites, or detected only by Swift or only by Fermi, respectively. 

\begin{figure}
    \centering
    \includegraphics[width=0.8\columnwidth]{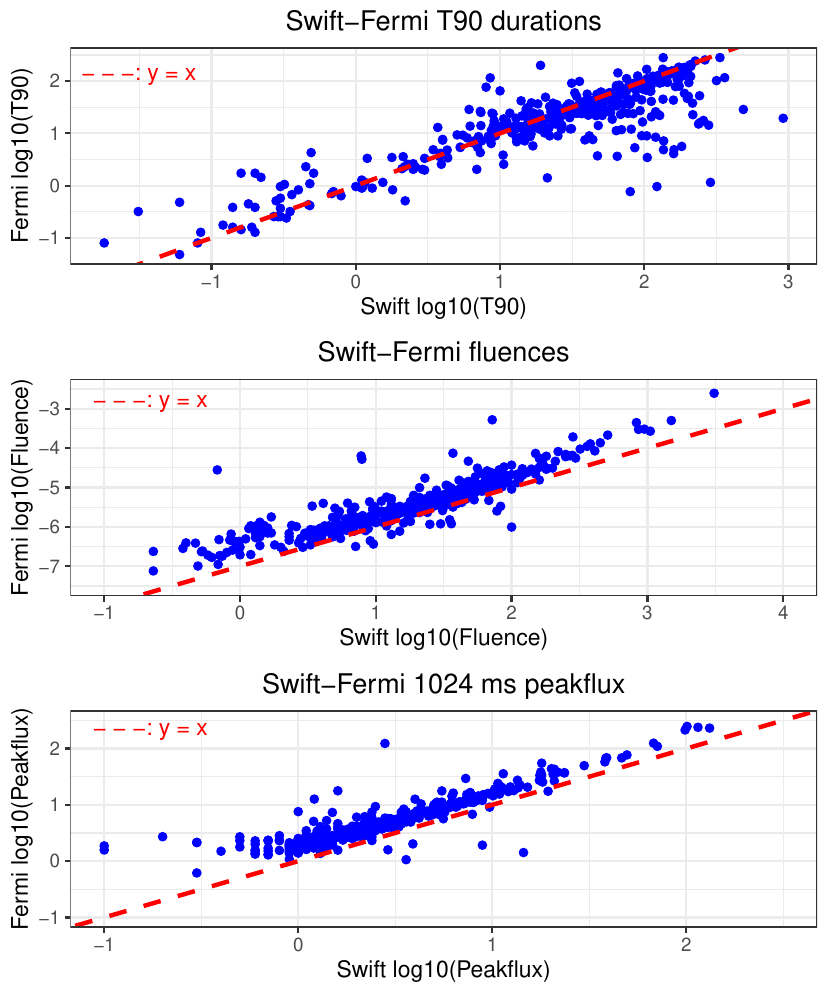}
    \caption{Comparison of T$_{90}$ [s] duration (top), fluence [$erg \, cm^{-2}$] (middle), and peak flux [$count \, cm^{-2} \, s^{-1}$] (bottom) of Fermi and Swift.The \textit{X} coordinate corresponds to the Swift and the \textit{Y} to the Fermi data. The energy range is $10-1000$ keV for Fermi GBM and $15-150$ keV for Swift BAT. At the medium T$_{90}$ the values obtained from the measurements of the two satellites are almost the same, while at the shorter and longer T$_{90}$ duration the values of Fermi and Swift are systematically higher, respectively. For fluence and peak flux the values obtained from Fermi measurements are systematically higher. The dashed red line indicates the same identical values obtained by the two satellites. }
    \label{fig:log-log-swift-fermi}
\end{figure}

Of course, only the time interval in which both Swift and Fermi were operating simultaneously should be taken into account in identifying the "widows". In identifying the "couples" this condition is fulfilled automatically. For the differences between the basic parameters of "couples" group observed by the different satellites see Fig.~\ref{fig:log-log-swift-fermi}.

To compare the GRBs, detected by the Swift and Fermi satellites, we used T$_{90}$ duration, fluence and peak flux,  physical parameters derived from the measurements of both satellites\footnote{For definitions of these parameters, see footnotes 1 and 2}. The parameters were determined from photons incoming in the $15-150$ keV energy range for the Swift \citep{2004ApJ...611.1005G} and $10-1000$ keV for the Fermi \citep{2009ApJ...702..791M}.

Fig.~\ref{fig:log-log-swift-fermi} shows even at first glance that the relationship between the quantities measured by BAT and GBM cannot be characterized simply by the $y = x$ line. For duration, the slope is different, while for fluence and peak flux, the values measured by Fermi are systematically higher. These differences can be explained by the fact that the energy range of GBM includes the energy range of BAT, however, it also detects photons with much higher energy, i.e. those that are already outside the energy range of BAT.

In the following, we study the differences in the values of the physical variables characterizing the "couples" and "widows" GRBs detected by BAT and GBM. The linear discriminant method was used for this purpose. We also study how the GRBs detected by both satellites differ in the observed variables. For this purpose, the canonical correlation was used. In both procedures, the linear (Pearson) correlation plays an important role. This type of correlation is sensitive to outliers in the data. A usual way using logarithmic variables to suppress their effect in the analysis. We proceeded in this way in our computations.  

\subsection{Linear discriminant analysis basics}

Linear discriminant analysis (LDA) is a method used in statistics to find a linear combination of features that characterizes or separates two or more classes of objects or events.

Let we have a set of $p$ measured variables on $n$ cases which are  assigned  to  one  of  the $k$ classes ($k=2$ in our case). We look for linear combination of the $\{x_1,x_2, . . .,x_p\}$ variables which give maximal separation between the groups of the cases. It means we are looking for the variable

\begin{equation}
y=n_1x_1+n_2x_2+ ... +n_px_p \quad \text{ where } \quad n_1^2+n_2^2+ ... + n_p^2=1
\end{equation}

\noindent with  a  suitable  chosen $\{n_1,n_2, . . .,n_p\}$ coefficients  ensuring a maximal separation between the classes.

\subsection{LDA of "couples" and "widows" in BAT and GBM data}

To get the best performing direction we performed linear discriminant analysis ($LDA$) in the parameter space \cite{LDA1,LDA2,LDA3,LDA4}. $LDA$ is available in the $MASS$ library of the \textbf{R} project \cite{MASS,racz_mnras_fermi_2018}. Performing $LDA$ on BAT data we got a very pronounced difference between the "couples" and "widows" GRBs detected by the Swift satellite.

Similarly to the analysis of Swift BAT data we can look for the most discriminating direction between the "couples" and "widows" in the parameter space of the observed Fermi GBM data.  The analysis demonstrated that the difference between "couples" and "widows" is much less pronounced than at GRBs detected by the Swift satellite. 

\subsection{Canonical correlation basics}

Canonical correlation (CC) assumes we have two set of variables: \textit{X} and \textit{Y}. The first set, \textit{X}, contains $\{x_1,x_2,...,x_m\}$ and \textit{Y}, the second one, $\{ y_1,y_2,...,y_r \} $ variables. We make $n$ observations for each variables. Using the linear combination of the $X$ and $Y$ variables we develop
\begin{equation}
U=(a_1)(x_1)+(a_2)(x_2)+...+(a_m)(x_m)
\end{equation}

and 
\begin{equation}
V=(b_1)(y_1)+(b_2)(y_2)+...+(b_r)(y_r) 
\end{equation}

\noindent asking: how can one select the "$a$" and "$b$" set of coefficients so that correlation between $U$ and $V$, obtained above, has the maximum value.

In our case at Swift and Fermi "couples" we have Swift (denoted with $X$) and Fermi (denoted with $Y$) data for the same GRBs, observed by both satellite. In this case $m=r=3$. The BAT and GBM data from the two set of variables represent the input of the canonical correlation.
For performing canonical correlations we used \textit{cc()} procedure in \textit{CCA} library of the \textbf{R} statistical package \cite{CCA}. We tested the significance of the variables obtained applying Wilks’ $\lambda$-test implemented in \textit{p.asym()} procedure in $CCP$ library of \textbf{R} \citep{CCP}.

\subsection{Canonical correlations between BAT and GBM "couples" data}

Maximizing the correlation between the $U$ and $V$ variables yields a unit vector ($a_1$, $a_2$, $a_3$, $a_4$) in the parameter spaces of the BAT variables and ($b_1$, $b_2$, $b_3$, $b_4$) in the parameter space of those in GBM. The vectors $\smash{\vec{a}}$ and $\smash{\vec{b}}$ denote the direction in the space of the BAT and GBM variables along which the correlation between the $U$ and $V$ is maximal. The components of the vectors $\smash{\vec{a}}$ and $\smash{\vec{b}}$, respectively, indicate how strongly the variables of BAT and GBM participate in this correlation. 

However, the direction thus obtained does not necessarily characterize all relationships between BAT and GBM variables. All directions perpendicular to directions $\smash{\vec{a}}$ and $\smash{\vec{b}}$ form a subspace in the parameter space of BAT and GBM, respectively, in which we can find another $\smash{(\vec{a},\vec{b})}$ pair, which denotes the directions along which the correlation between $U$ and $V$ is maximal. Repeating this procedure, we get the variables $(U1, U2, U3)$ and $(V1, V2, V3)$ in the BAT and GBM spaces, respectively. The components of their $\smash{\vec{a}}$ and $\smash{\vec{b}}$ vectors indicate the physical variables of GRBs. The whole process is coded in the \textit{cc()} procedure. However, the correlation between the $U$ and $V$ variables thus obtained is not necessarily significant. Significance is obtained from the \textit{p.asym()} procedure.

\section{Discussion}

\subsection{Remarks to LDA on Swift BAT and Fermi GBM data}

\begin{figure}
    \centering
    \includegraphics[width=\columnwidth]{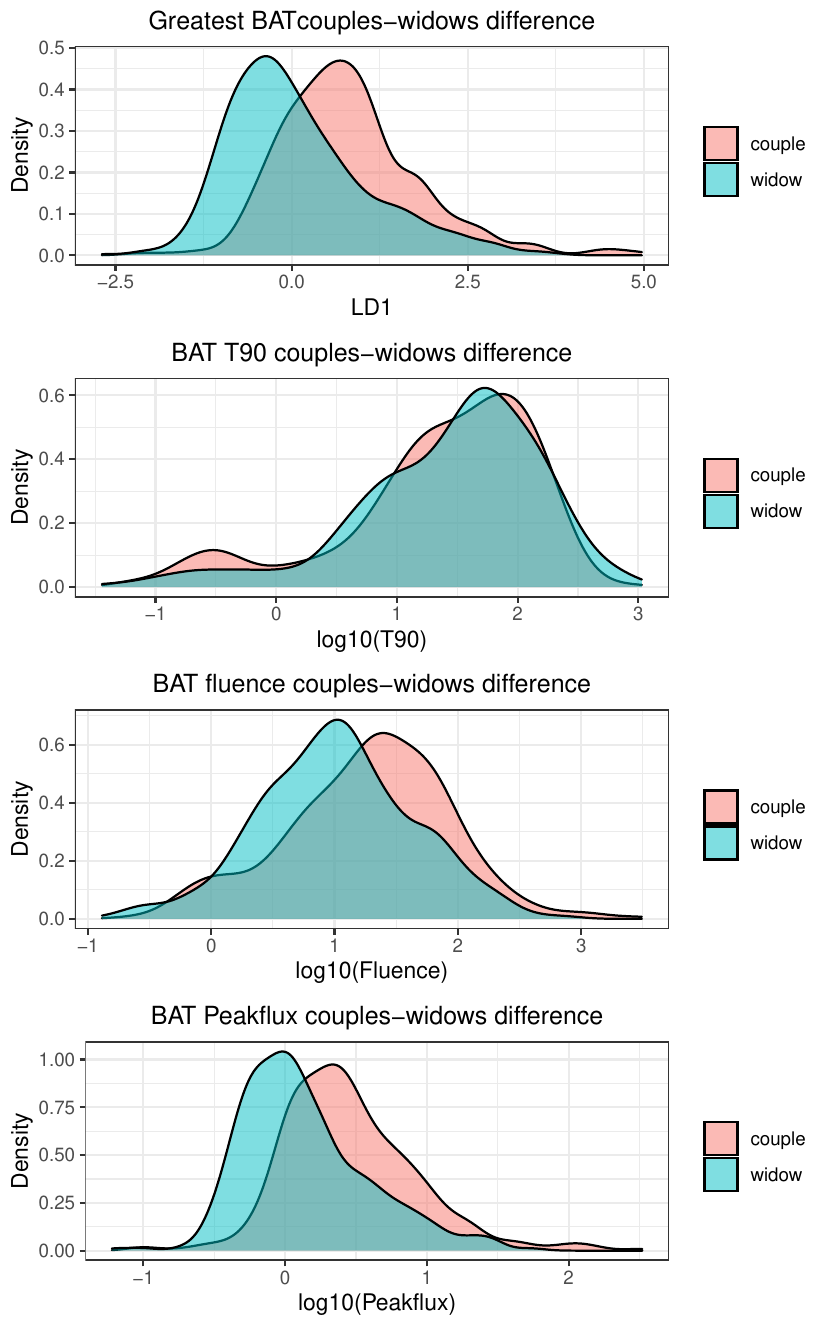}
    \caption{Separation of Swift BAT "couples" and "widows" GRBs along the best discriminating direction (LD1) obtained by the LDA (upper panel), and the degree of differences of each measured variable. Apparently, the highest difference in measured variables is given by the peak flux.  See text for the definition of the LD1 dimensionless variable.}
    \label{fig:swift-LD1}
\end{figure}

The discriminant analysis between Swift "couples" and "widows" revealed that the  the joint distribution of the Swift "couples" and "widows" T$_{90}$, fluence and peak flux variables differ at a very high level of significance (Fig. \ref{fig:swift-LD1}).
The LD1 variable describing the highest discrimination between Swift "couples" and "widows" has the highest correlation with peak flux and closely followed by fluence. The highest contribution, correlation to the LD1 discriminant variable is given in absolute value by the peak flux (0.95), followed by the fluence (0.56), and the T$_{90}$ duration at the end (0.15). (The difference between the  mean values in "couples" and "widows" groups is given in Table \ref{batlda}).

\begin{table*}

\begin{minipage}{0.49\textwidth}
\centering
\begin{tabular}{rlrccc}
  \hline
 & Group & LD1 & $\log_{10}(T_{90})$ & $\log_{10}(Flu)$ & $\log_{10}(Peak)$ \\ 
  \hline
1 & couples & $-0.82$ & 1.35 & 1.27 & 0.46 \\ 
  2 & widows & $-0.08$ & 1.43 & 1.02 & 0.14 \\ 
   \hline
\end{tabular}
\caption{Differences  between "couples" and "widows" groups in BAT LDA. The error probability for rejecting the null hypothesis, i.e. differing the groups only by chance, is less than $2\cdot 10^{-16}$.} 
\label{batlda}
\end{minipage}\hfill
\begin{minipage}{0.49\textwidth}
\centering
\begin{tabular}{rlrccc}
  \hline
 & Group & LD1 & $\log_{10}(T_{90})$ & $\log_{10}(Flu)$ & $\log_{10}(Peak)$ \\ 
  \hline
1 & couples & 2.86 & 1.22 & $-5.43$ & 0.72 \\ 
  2 & widows & 3.07 & 1.10 & $-5.57$ & 0.64 \\ 
   \hline
\end{tabular}
\caption{Differences  between "couples" and "widows" groups in GBM LDA. The probability for differing the group only by chance is less than $3.23 \cdot 10^{-5}$. It is still significant, but much less pronounced than Swift BAT.} 
\label{gbmlda}
\end{minipage}

\end{table*}

The mean values of these variables are higher in the "couples" than in the "widows" group.
The means of  T$_{90}$ duration are higher in the "widows" group. The smaller mean value of T$_{90}$ is caused by a slight surplus of short GRBs in the "couples" group.

An interesting result of the LDA is an apparent deficit of intermediate duration GRBs in the T$_{90}$ distribution at "couples", and in the contrary, the short duration GRBs are somewhat fewer at the "widows" (Fig. \ref{fig:swift-LD1}).

These results are consistent with that obtained by \citet{2016ApJ...818..110B} finding that BAT detects weaker short duration GRBs than GBM.

 In the case of BAT, the largest difference between "couples" and "widows" is in peak flux, but there is also a significant difference in fluence values.

\begin{figure}
    \centering
    \includegraphics[width=\columnwidth]{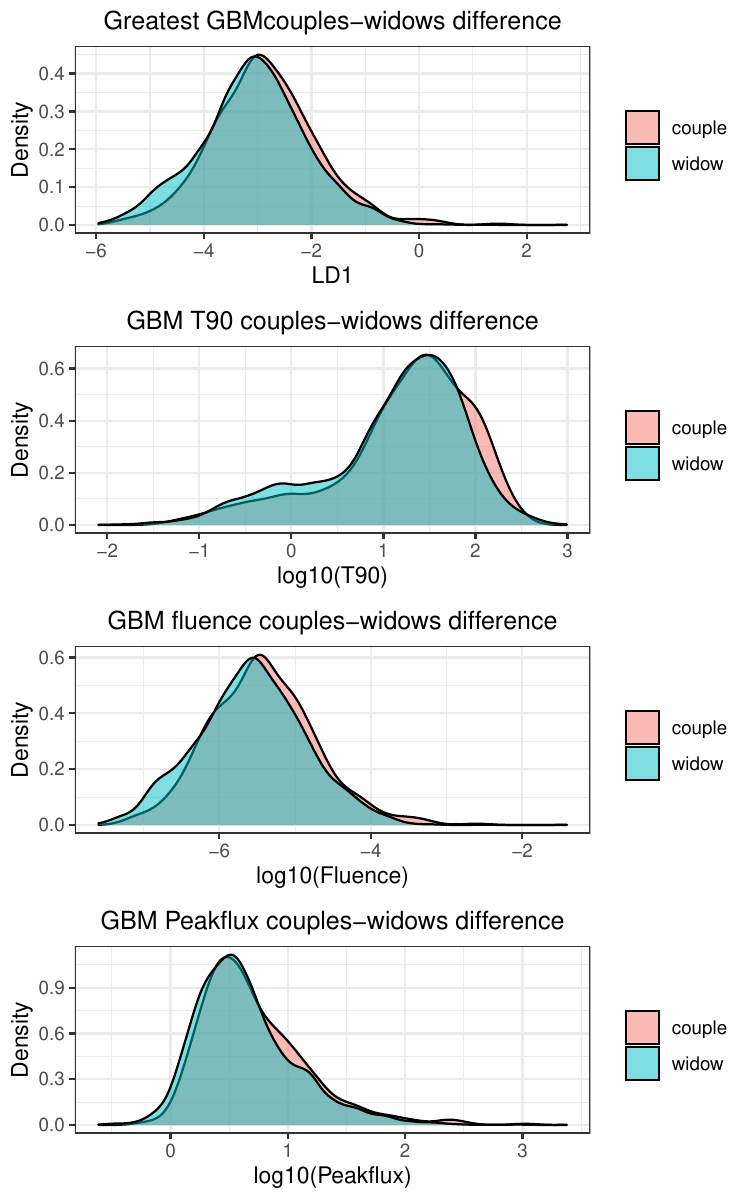}
    \caption{Separation of Fermi GBM "couples" and "widows" GRBs along the best discriminating direction (LD1) obtained by the LDA (upper panel) and the degree of contribution of each measured variable. The units of variables are as before. Apparently, the difference between "couples" and "widows" is much less pronounced than in BAT. The greatest contribution to the "couples" "widows" difference  is given by the fluence }
    \label{fig:fermi-LD1}
\end{figure}

Apparently, the distribution of the Fermi GBM "couples" and "widows" variables (Table \ref{gbmlda}, Fig. \ref{fig:fermi-LD1}) differs much less from that of the Swift BAT. This phenomenon may be partly explained by the fact that the Swift sees a much smaller part of the sky compared to the Fermi. So at some given event, there could be also GRBs among the Fermi "widows" category that would belong to the "couples" group if they fell into Swift’s field of view.

In case of GBM, the most significant difference appears in the distribution of fluences and durations. 
The highest contribution, correlation to the LD1 discriminant variable is given in absolute value by the fluence (0.86), followed by the T$_{90}$ (0.85) and the peak flux (0.62), at the  end. 

The duration of the "couples" bursts appears to be significantly longer. As the longer duration bursts are softer, a higher percentage of incoming photons fall within the range of energy detected by BAT. The "couples" bursts' fluence is also larger than the "widows" due to the correlation with duration.  

\subsection{Remarks to canonical correlation between BAT and GBM "couples"}{\label{cc}}

\begin{figure}
    \centering
    \includegraphics[width=\columnwidth]{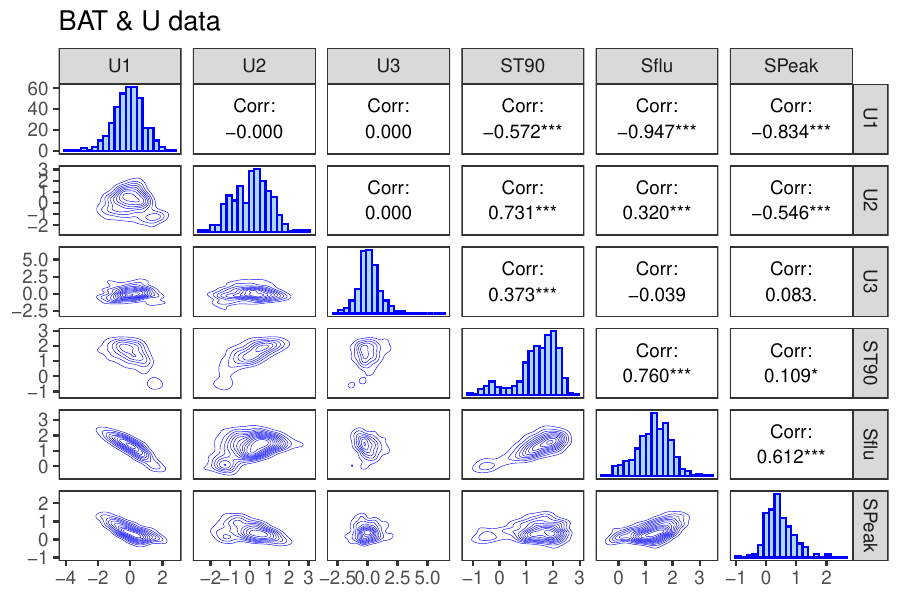}
    \caption{Matrix plot of BAT data and the canonical variables obtained from BAT (\textit{U1,U2,U3}). Lower panel shows 2D densities the upper one the correlations between the variables. The significance level of the correlation indicated with stars at the right side of the numbers. Seemingly, the first, the strongest, canonical variable (\textit{U1}) has the tightest correlation with fluence.}
    \label{fig:bat-u}
\end{figure}

\begin{figure}
    \centering
    \includegraphics[width=\columnwidth]{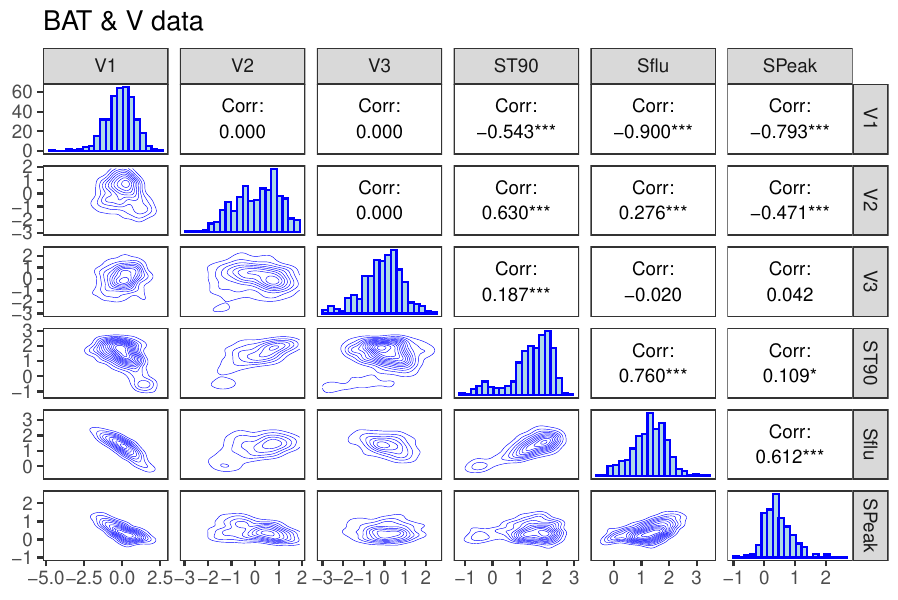}
    \caption{Matrix plot of BAT data and the canonical variables obtained from GBM ($V1,V2,V3$). Lower panel shows 2D densities the upper one the correlations between the variables. The significance level of the correlation indicated with stars at the right side of the numbers. Seemingly, the first, the strongest, canonical variable (\textit{V1}) has the tightest correlation with fluence. }
    \label{fig:bat-v}
\end{figure}

\begin{figure}
    \centering
    \includegraphics[width=\columnwidth]{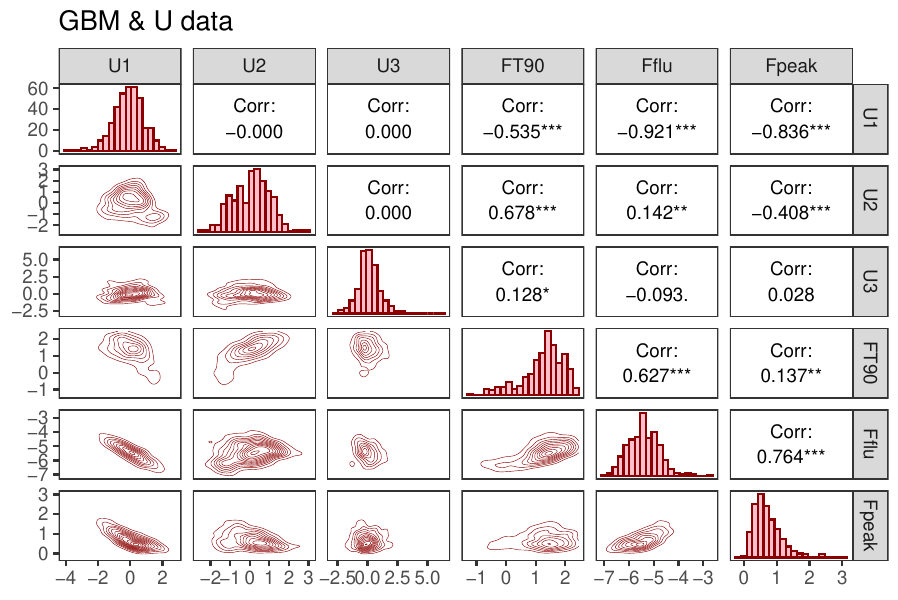}
    \caption{Matrix plot of GBM data and the canonical variables obtained from BAT (\textit{U1,U2,U3}). Lower panel shows 2D densities the upper one the correlations between the variables. The significance level of the correlation indicated with stars at the right side of the numbers. Seemingly, the first, the strongest, canonical variable (\textit{U1}) has the tightest correlation with fluence. }
    \label{fig:gbm-u}
\end{figure}

\begin{figure}
    \centering
    \includegraphics[width=\columnwidth]{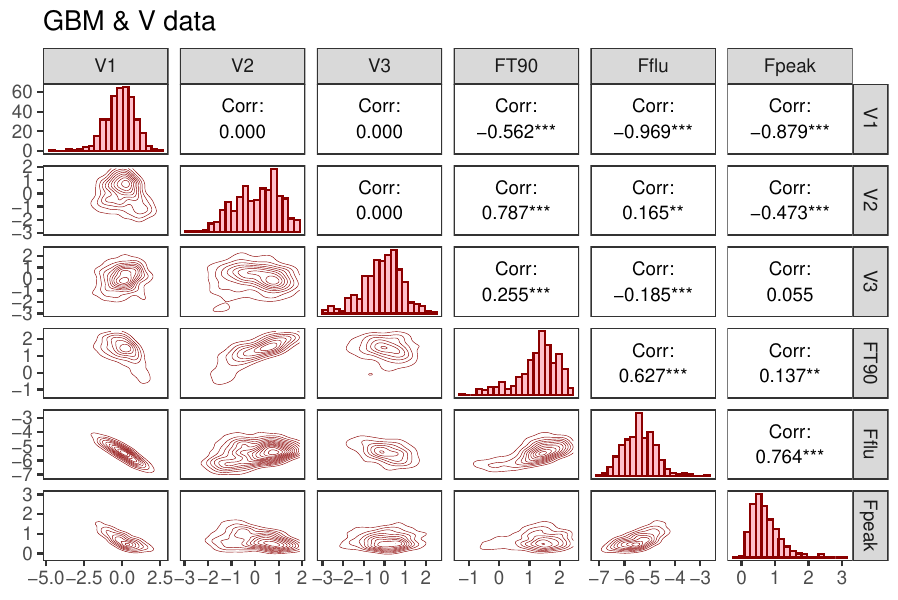}
    \caption{Matrix plot of GBM  data and the canonical variables obtained from GBM (\textit{V1,V2,V3}). Lower panel shows 2D densities the upper one the correlations between the variables. The significance level of the correlation indicated with stars at the right side of the numbers, Seemingly, the first, the strongest, canonical variable (\textit{V1}) has the tightest correlation with fluence.}
    \label{fig:gbm-v}
\end{figure}

Using the canonical variables obtained in the analysis we computed their correlations (canonical loadings) with the original ones. Canonical correlations resulted in tree canonical variables representing significant relationships between BAT and GBM data. The results of canonical correlation are summarized in Figs. \ref{fig:bat-u} and \ref{fig:bat-v} for the BAT and Figs. \ref{fig:gbm-u} and \ref{fig:gbm-v} for the GBM variables.

\begin{figure}
    \centering
    \includegraphics[width=\columnwidth]{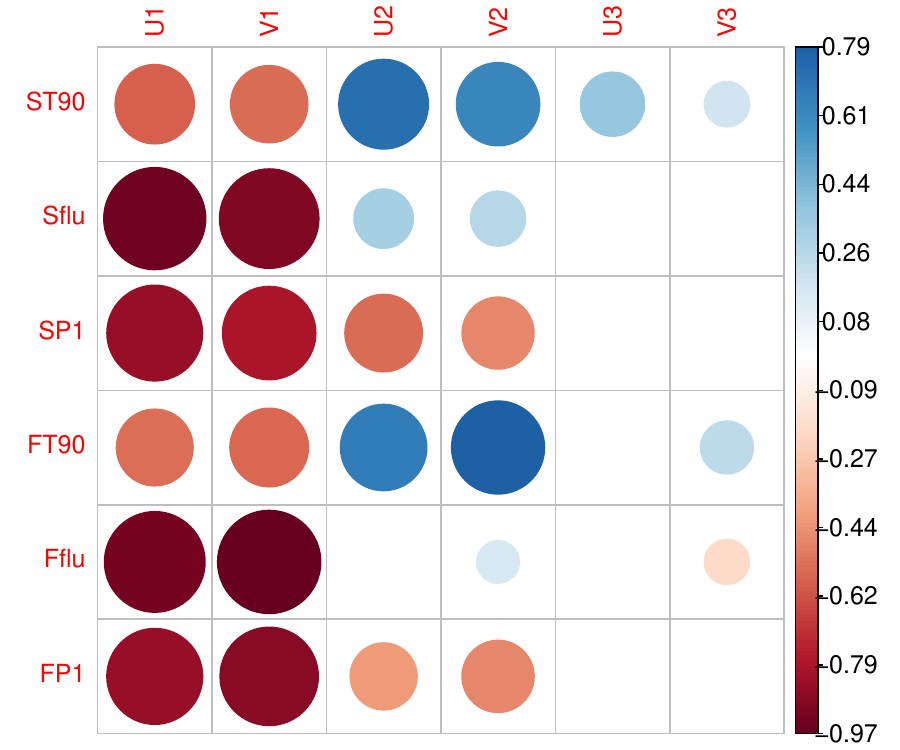}
    \caption{Coloured cross correlation between the BAT and GBM data with the significant canonical variables. The size and colour of the circles indicate the strength of correlation. Positive correlations marked with blue, negative with red colour. Correlations not reaching $3\sigma$ significance are left blank. The strongest canonical variable pair (\textit{U1,V1}) has the tightest relation to fluence for both BAT and GBM.
    }
    \label{fig:ccorr}
\end{figure}

The strongest (\textit{U,V}) pair (\textit{U1,V1}) dominated by the fluences in both of the Swift and Fermi data. Since T$_{90}$ and peak flux are correlating with fluence they also have strong correlations with the (\textit{U1,V1}) pair.

Both of them strongly correlate with the pair ($U2,V2$). Since the canonical variables are perpendicular to each other, this does not result from a correlation with fluence, but from a direct relationship between BAT and GBM duration and peak flux.

The third ($U3,V3$) canonical variables show a weak but significant relationship between the BAT and the GBM durations. As Figs \ref{fig:bat-u},\ref{fig:bat-v},\ref{fig:gbm-u},   \ref{fig:gbm-v} and \ref{fig:ccorr} demonstrates both BAT and GBM durations has some but decreasing level of correlations with all the canonical variables.  
 
 The GBM bursts average peak energy is around 200 keV which is outside the sensitivity range of BAT \citep{2015AdAst2015E..22P}. Therefore, a significant fraction of photons detected and used in GBM durations is not detected by BAT may causing a nonlinear relationship between BAT and GBM durations. Canonicial correlation is a linear theory and therefore requires a system of more orthogonal vector for accounting nonlinear relationships

\section{Classification of Swift and Fermi GRBs}

\begin{figure}
    \centering
    \includegraphics[width=\columnwidth]{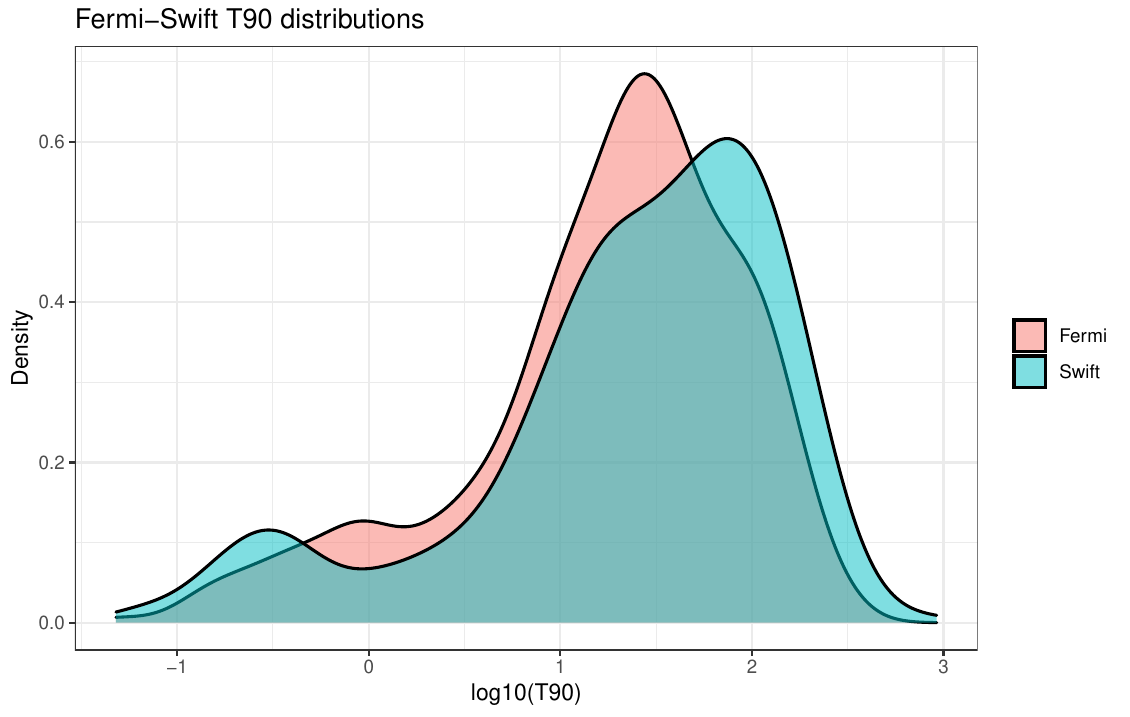}
    \caption{Distribution of $T_{90}$ duration measured by Fermi (light red) an Swift (light cyan) satellites. The durations measured by Fermi are more concentrated at medium values, while those measured by Swift are more concentrated at short and long values, respectively.}
    \label{fig:t90durdist}
\end{figure}

According to Fig. \ref{fig:t90durdist}, the duration of bursts detected jointly by Fermi and Swift is systematically longer based on Fermi measurements than the short ones, but the opposite is true for the long ones. If the durations obtained from the measurements of the two satellites were the same, the distribution in Fig. \ref{fig:log-log-swift-fermi} could be fitted with a line with a slope of 1. However, the slope of the line that best fits the points is $0.746 \pm 0.025$ at more than $5 \sigma$ significance level.

We also mentioned in the introduction that burst triggering procedure and the spectral range of detection are different for the two satellites. Since the two satellites see the same burst, the actual physical duration of the phenomenon must be the same. However, changes in the physical parameters of the outburst as a function of time occur differently due to the different technical design of BAT and GBM. \citep[See][to determine the duration of burst for BAT]{2016ApJ...829....7L,2020ApJ...893...46V}.

Short bursts are generally harder, so they trigger GBM earlier and stay longer above detection level. For the long ones, since they are softer, it's just the opposite in particular at the last stage of their spectral evolution. This is reflected in the deviation of points from $y=x$ line seen in Fig. \ref{fig:log-log-swift-fermi} displaying the duration of the jointly observed bursts.

\subsection{Fitting T$_{90}$ distributions of GRBs jointly detected by Fermi and Swift} 

\begin{figure}
     \centering
    \includegraphics[width=\columnwidth]{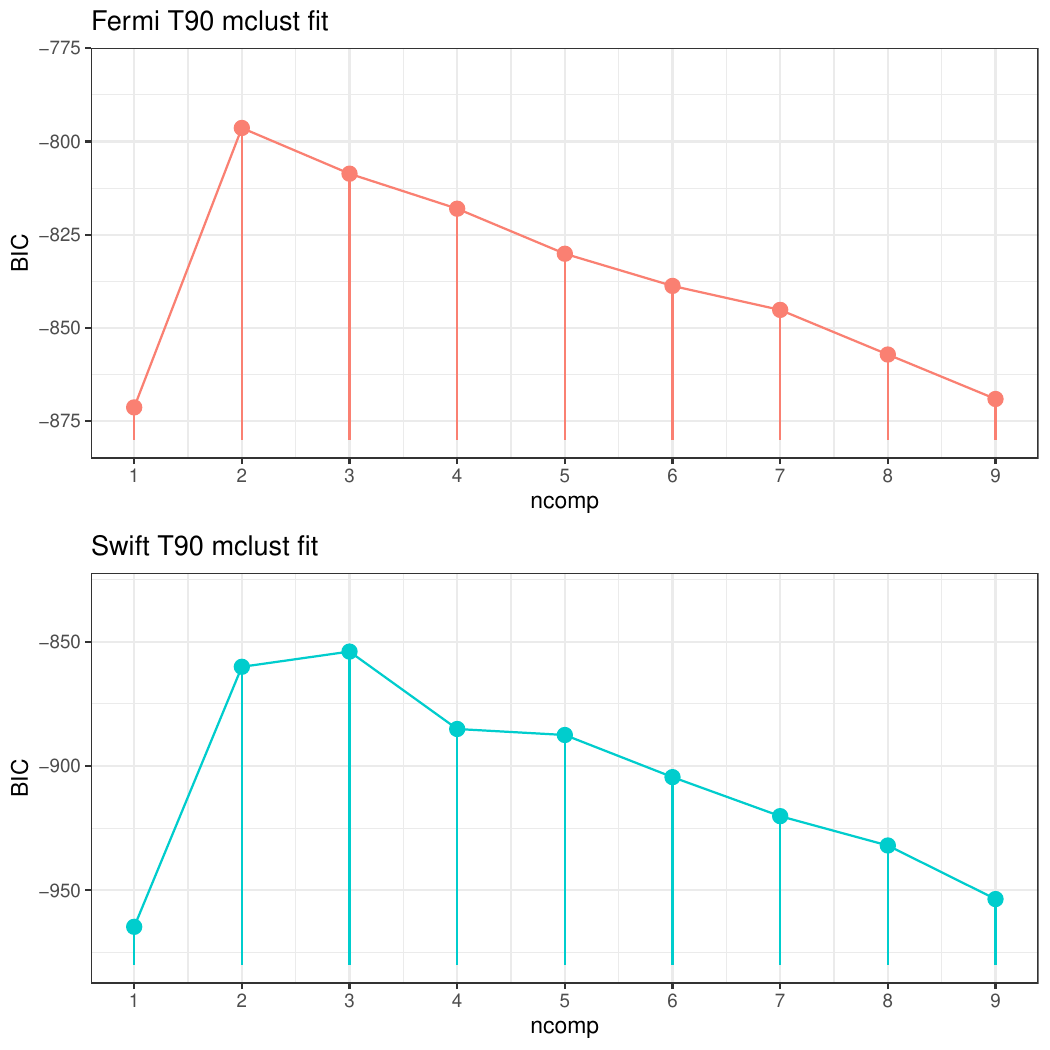}
    \caption{BIC values of the fitted multi-component lognormal models. The maximum value of BIC for Fermi (ligh tred) is at  two components, while for Swift (cyan) it is three components, although both satellites measured the same GRBs. }
    \label{fig:components}
\end{figure}

To perform fitting the T$_{90}$ distribution by means of superposing lognormal distributions we used \textit{Mclust()} procedure in \textit{mclust} library of \textbf{R}.  Using these  lognormal mixture models the procedure computes BIC values\footnote{Bayesian information
criterion value is formally defined as $BIC = k\cdot ~ln(n)-2\cdot ln(L)$, where k is the number of
parameters estimated by the model, n the number of data points and L the maximized value of
the likelihood function} starting with $k=1$ component and proceeds to a given higher k value. The optimum $k$ number of components is obtained at giving the highest BIC value. The result is given in Fig. \ref{fig:components} for Fermi (red  colour) and Swift (cyan colour), respectively.

We found that the number of its best-fit distribution components was different for Fermi and Swift measurements, although in both cases the GRBs were the same. It is worth mentioning \citet{2022GalaxSalmon} made a 
two dimensional clustering of Swift/BAT and Fermi/GBM Gamma-ray Bursts and also found two groups for  GBM and three for  BAT. 

As we mentioned above, in Fig. \ref{fig:t90durdist} the Swift is stronger on the edge of the T$_{90}$ range, while the Fermi in the middle. Since  both distributions are given by the same GRBs, we have to conclude that the T$_{90}$ distribution obtained from the observations cannot be inferred directly for the number of physical engine types operating in the background.

As we pointed out, the effect can be explained by considering the different energy sensitivity ranges used by Fermi and Swift satellites to calculate the physical parameters.  As we mentioned,  the Fermi parameters are calculated from the photons in the energy range of $10-1000$ keV and that of Swift in the $15-150$ keV range.
Since bursts are initially harder and then gradually become softer during eruption (see, e.g. \cite{racz_2018an}) Fermi may notice them earlier than Swift. Although, the $15-150$ keV range is detected by both satellites, but here Swift is more sensitive. Therefore, bursts can be followed for a longer time period.

\section{Conclusions}

We examined how the technical properties of the Swift and Fermi satellites affect the observable properties of the GRBs they detect. In our study, we examined the data obtained from the Swift BAT and Fermi GBM instruments. These data were T$_{90}$, fluence, peak flux for BAT and T$_{90}$, fluence, peak flux for Fermi GRBs.

In order to identify GRBs detected jointly by Swift and Fermi we looked for coincidences in GRB angular position - trigger time parameter space of both satellites. For this purpose we used the \textit{knn()} procedure available in \textit{FNN} library of the \textbf{R} statistical package.

Based in these identifications we separated the "couples" and "widows" GRBs, detected simultaneously by both satellites and only by one of them.
In case of the "couples" the values of T$_{90}$ are satisfactorily the same for the medium duration, while the data of the Fermi GBM are systematically higher in the case of the short ones and the data of the Swift BAT in the case of the long ones.
For fluence and peak flux, the Fermi satellite measured a systematically larger value for the same GRB.

Using the linear discriminant analysis (LDA) we compared the physical properties of "couples" and "widows" GRBs in BAT and GBM. For this purposed we utilised the \textit{lda()} procedure available in \textit{MASS} library of \textbf{R} statistical package. LDA resulted a direction in the parameter space of observed variables along with the difference between the "couples" and "widows" group is the greatest. We obtained that peak flux has the highest discriminant power in case of Swift and fluence in Fermi.

Using canonical correlation we studied the strength of the relationship between GRB parameters measured by Swift and Fermi, respectively. This relationship is represented by three orthogonal canonical variable pairs. The strongest of these is the largest contribution from fluence for both Swift and Fermi. 

We tested the hypothesis that the number of lognormal distributions used to fit GRBs to T$_{90}$ distribution could be inferred for the physical mechanisms responsible for eruptions. For this purpose, we compared the distributions of the T$_{90}$ jointly detected by the two satellites in the Swift and Fermi data, separately. Since the GRBs used for this analysis are the same at both satellites one expect the same number of lognormal components necessary to fit the T$_{90}$ distributions.

In contrast, we obtained that the number of lognormal components required is \textit{three} for Swift, while it is only \textit{two} for Fermi. Since the GRBs used for the analysis were the same in both cases, we concluded that it is not possible to infer the number of physical mechanisms responsible for GRBs from the T$_{90}$ distribution alone.

\section*{Acknowledgements}

The authors would like to thank Jakub {\v{R}}ípa and Mariusz Tarnopolski for numerous discussions on GRBs. We are indebted to Jakub {\v{R}}ípa for initiating the writing of this paper. The authors thank the Hungarian TKP2021-NVA-16 and
OTKA K-146092 program for their support.






\bibliographystyle{mnras}
\bibliography{main,lajos,horv22b}

\bsp	
\label{lastpage}

\end{document}